\documentclass[epj]{svjour}
\pdfoutput=1
\usepackage{graphics}
\usepackage{graphicx} 
\usepackage{color}

\usepackage{xr}
\makeatletter
\newcommand*{\addFileDependency}[1]{
	\typeout{(#1)}
	\@addtofilelist{#1}
	\IfFileExists{#1}{}{\typeout{No file #1.}}
}
\makeatother

\newcommand*{\myexternaldocument}[1]{
	\externaldocument{#1}
	\addFileDependency{#1.tex}
	\addFileDependency{#1.aux}
}

\myexternaldocument{supporting}
\begin{document}
\title{Microfluidic device coupled with total internal reflection microscopy for in situ observation of precipitation}
\author{Jia Meng\inst{1} \and Jae Bem You\inst{1} \and Gilmar F. Arends\inst{1} \and Hao Hao\inst{2} \and Xiaoli Tan\inst{1}\thanks{\emph{}xiaolit@ualberta.ca} and Xuehua Zhang\inst{1}\thanks{\emph{}xuehua.zhang@ualberta.ca}%
}                     
%
%
\institute{Department of Chemical and Materials Engineering, University of Alberta, Alberta, Canada T6G 1H9 \and Central Faculty Office (FSET), Swinburne University, Melbourne, Australia 3122.}
\date{Received: date / Revised version: date}
%
\abstract{
In situ observation of precipitation or phase separation induced by solvent addition is important in studying its dynamics. Combined with optical and fluorescence microscopy, microfluidic devices have been leveraged in studying the phase separation in various materials including biominerals, nanoparticles, and inorganic crystals. However, strong scattering from the subphases in the mixture is problematic for in situ study of phase separation with high temporal and spatial resolution. In this work, we present a quasi-2D microfluidic device combined with total internal reflection microscopy as an approach for in situ observation of phase separation. The quasi-2D microfluidic device comprises of a shallow main channel and a deep side channel. Mixing between a solution in the main channel (solution A) and another solution (solution B) in the side channel is predominantly driven by diffusion due to high fluid resistance from the shallow height of the main channel, which is confirmed using fluorescence microscopy. Moreover, relying on diffusive mixing, we can control the composition of the mixture in the main channel by tuning the composition of solution B. We demonstrate the application of our method for in situ observation of asphaltene precipitation and $\beta$-alanine crystallization.} 
\maketitle
\section{Introduction}
\label{sec:1}
In situ observation and control of phase separation or precipitation process is important in many areas including growth of nanoparticles,\cite{wang2020,harada2012} crystallization,\cite{xuan2019,driessche2010} and biomineralization.\cite{wang2012} A representative phase separation occurs when the concentration of a solute dissolved in a solvent reaches a critical concentration by the addition of an antisolvent. At the critical concentration, the solute is oversaturated, and the mixture is spontaneously separated into two subphases. Depending on the type of the solute, solid phase formation may be formed as precipitates or crystals. In all cases, both the chemical composition of the final mixture (i.e. solute + solvent + antisolvent) and controlled mixing between the solution (i.e. solute + solvent) and antisolvent determine the phase separation outcome. Precipitation is of particular interest during paraffinic froth treatment in oil sand extraction process, where asphaltene precipitates are formed from the addition of antisolvent.\cite{xu2018}

Recently, microfluidic devices have been widely used for phase separation of various materials ranging from crystals,\cite{gong2015,kim2017} drug nanoparticles,\cite{karnik2008} and lipid nanoparticles.\cite{zhigaltsev2012} Phase separation in certain microfluidic devices can be predominantly driven by diffusive mixing between the solute + solvent mixture and the antisolvent. By leveraging control of diffusion, the influence of chemical composition and mixing on the phase separation dynamics can be decoupled. Moreover, the outcome of phase separation can be visualized in situ via optical or fluorescence microscopy.  For instance, Sekine \textit{et al.} was able to observe halite crystal growth in situ using a microfluidic channel.\cite{sekine2011} Desportes \textit{et al.} used fluorescence microscopy coupled with a microfluidic set-up to observe the nanocrystallization of rubrene.\cite{desportes2007} However, the spatiotemporal resolutions of optical and fluorescence microscopy are not high enough for detailed study of the phase separation dynamics in early stage. Furthermore, in some cases, an in situ precipitation study is complicated by the strong absorption of light by the dark medium, for example, asphaltene precipitation from toluene solution.

Coupling the advantages of microfluidics with total internal reflection fluorescence (TIRF) microscopy provides an attractive solution to study the phase separation precipitation dynamics. TIRF is an imaging method that uses the evanescent wave produced at an interface between two media with different refractive index. As the penetration depth of evanescent field is shallow ($<$ 1 $\mu m$), combining TIRF and microchamber is a powerful method to study any change that occurs near the chamber surface. For instance, TIRF has been applied to investigate the liquid-liquid phase separation during mixing process\cite{zhang2015}, early stage diffusive growth of droplets,\cite{dyett2018} and real-time chemical reactions in surface nanodroplets.\cite{dyett2020}

In this work, we demonstrate a microfluidic method combined with TIRF for in situ observation of precipitation and phase separation. A bilayer quasi-2D microfluidic device – composed of a shallow main channel and a deeper side channel – is constructed to enable controlled diffusive mixing between two solutions to drive the phase separation. Using fluorescence microscopy, we validate the diffusive mixing process between solutions in the observation area in the main channel. By characterizing fluorescence intensity, we also quantify the change of chemical composition in space and in time in the observation area. We further demonstrate two examples of phase separation using the proposed setup, namely precipitation of asphaltene and crystallization of $\beta$-alanine. 

\section{Experimental}
\label{sec:2}
\subsection{Device fabrication}
\label{subsec:2.1} 
\begin{figure*}
	\centering
	\includegraphics{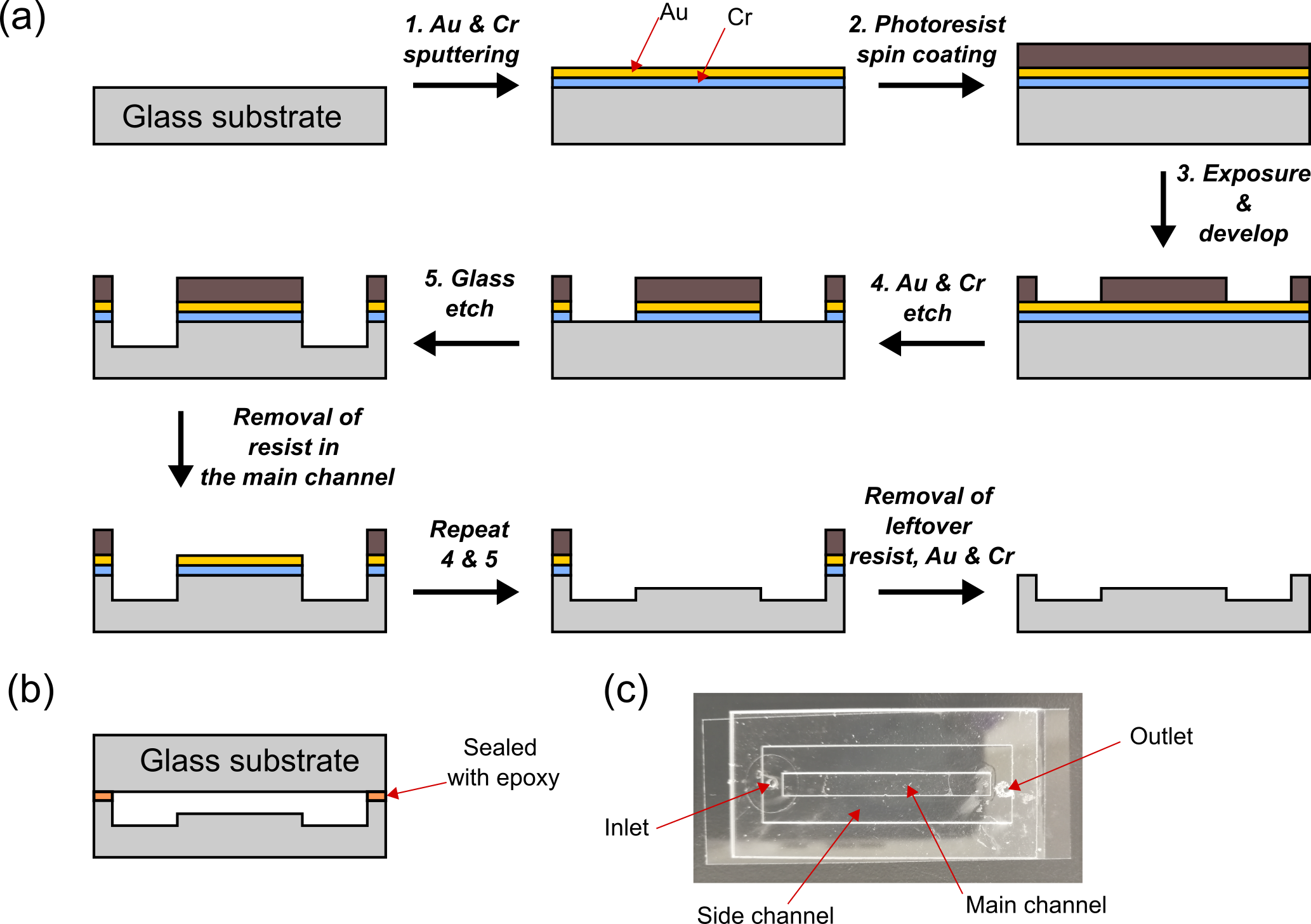}
	\caption{a) Schematic showing photolithographic procedure for fabricating the channel on glass substrate. b) The patterned glass substrate is sealed with a blank glass substrate using epoxy. c) Photograph of the fabricated microfluidic chip}
	\label{fig.1}       
\end{figure*}

\begin{figure*}
	\centering
	\includegraphics{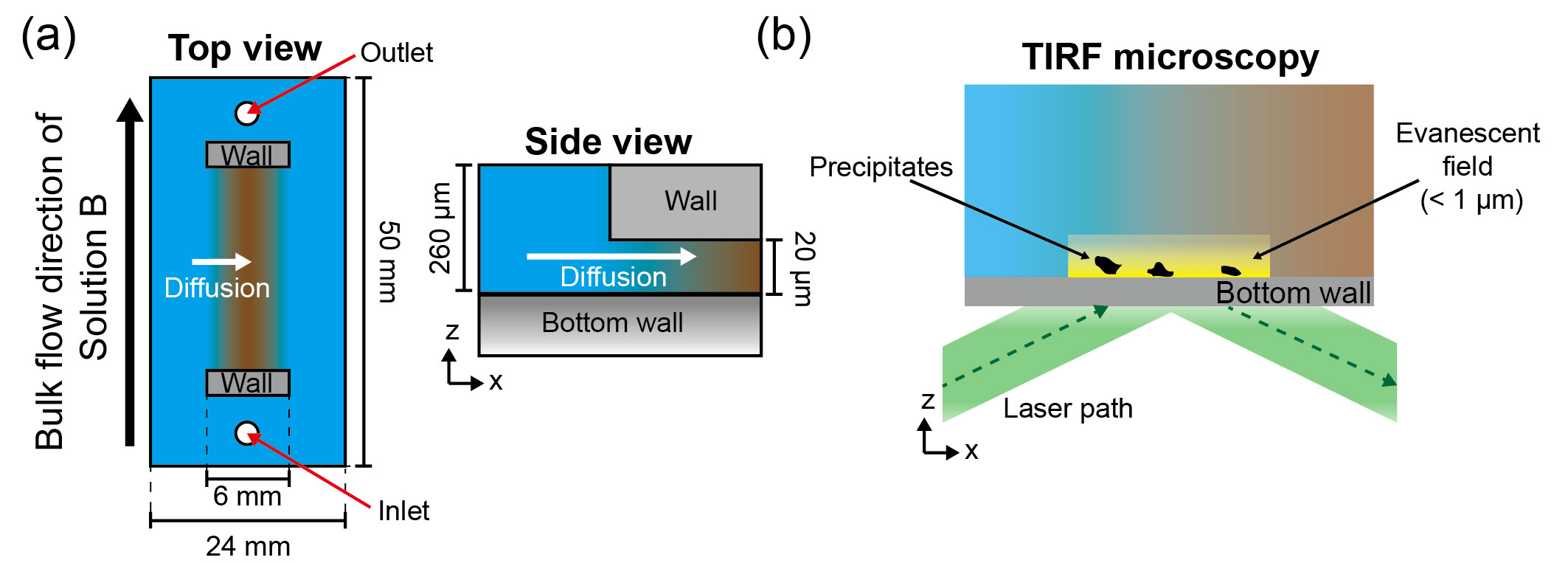}
	\caption{a) Top view and side view of the fluid chamber. The chamber consists deep side channel with 260 $\mu m$ depth (blue) and shallow main channel with 20 $\mu m$ depth (brown). b) Mechanism of total internal reflection fluorescence (TIRF) microscope. Only the asphaltene in evanescent flied (yellow) can be detected.}
	\label{fig.2}       
\end{figure*}

The quasi-2D microfluidic device was fabricated via standard photolithography process followed by wet etching as shown in Fig. \ref{fig.1}. Borofloat glass wafer (81\% SiO$_{2}$, 13\% B$_{2}$O$_{2}$, 4\% Na$_{2}$O/K$_{2}$O, 2\% Al$_{2}$O$_{3}$. Borofloat wafer was first cleaned with piranha solution (3 : 1 = H$_{2}$SO$_{4}$ : H$_{2}$O$_{2}$) and rinsed by de-ionized water. Chrome (20 $nm$) and gold (150 $nm$) were sputtered on the surface of the Borofloat sequentially as masking layer. Photoresist (AZ1529) was spin coated on the metal-masked substrate, which was baked for 1 $min$. The substrate was exposed under UV light through a photo mask with pre-designed side channel pattern on it. The exposed patterned photoresist was removed by developer. Gold and chrome layers were then etched by gold and chrome etchant. The Borofloat substrate was then etched by hydrogen fluoride solution to create 240 $\mu m$ side channel. After etching of side channel, blue tape was used to protect both ends of main channel. Photoresist on main channel was then manually removed by acetone wetted cotton swab. Gold, chrome and Borofloat etching steps were repeated to get 20 $\mu m$ etch of the main channel. Borofloat was cleaned by de-ionized water and the remaining gold and chrome layer were stripped by etchants. The patterned Borofloat was then bonded with a glass cover slip (24 $mm$ $\times$ 50 $mm$ $\times$ 170 $\pm$ 5 $\mu m$, Azer Scientific) by epoxy. Similar to the structure of microchamber used in our previous work,\cite{lu2017} the detailed dimensions of the quasi-2D microfluidic device are shown in Fig. \ref{fig.2}a. The narrow main channel is ~ 20 $\mu m$ in depth and 6 $mm$ in width, flanked with two deep side channels with ~ 260 $\mu m$ in depth and 3 $mm$ in width. The lengths of both the main channel and side channel are about 3 $cm$.

\subsection{Visualization of diffusive mixing using the quasi-2D microfluidic device}
\label{subsec:2.2}
All of the experiments were conducted at room temperature ($\sim$19 - 21 $^{o}C$). To confirm the diffusive mixing inside the microchamber, the mixing process between toluene and n-pentane was followed by using a fluorescence microscope (Nikon ECLIPSE Ni) equipped with a camera (Nikon DS-Fi3). Toluene (ACS grade, Fisher Scientific, 99.9+\%), doped with a trace amount of Nile Red (Fisher Scientific), was injected to fill the entire chamber. Then n-pentane (Fisher Scientific, 98\%) was injected from one end of the side channel at a flow rate 5 $mL/h$ by a syringe pump (NE-1000, Pumpsystems Inc.) as sketched in Fig. \ref{fig.2}. As flowing in the side channel along y-axis, n-pentane diffused transversely into the narrow quasi-2D main chamber to mix with toluene along x-axis. The location of the field of view was set at a distance of 80 $\mu m$ from the side channel through the transparent top glass of the main chamber.

Green laser (559 $nm$) was used to excite Nile Red and the emission was monitored at 635 $nm$. The change in the fluorescence intensity of the dye in the liquid was recorded at 2 frames per second by the fluorescence microscope (Nikon ECLIPSE Ni) equipped with a camera (Nikon DS-Fi3). The fluorescence intensity was converted to gray scale value and read using MATLAB software (The MathWorks Inc.). The fluorescence intensity as a function of time is governed by the mixing process, which is vital dataset to confirm whether diffusion mixing is dominant and to estimate the chemical composition of solution at given location.

\subsection{In situ observation of precipitation and phase separation using the quasi-2D microfluidic chamber}
\label{subsec:2.3}
The microfluidic device was initially filled with asphaltene solution in toluene (solution A). Then, solution B – n-pentane : toluene mixture at various ratios – was injected through the deep side channel with flow rate 5 $mL/h$ by a syringe pump where n-pentane is an antisolvent of asphaltene. Part of solution B diffuses transversely into the main channel and mixes with solution A. Asphaltene gradually precipitated with n-pentane diffusion in the main channel and can be detected by DeltaVision OMX Super-resolution microscope (TIRF 60$\times$/1.49 NA objective lens) (GE Healthcare UK limited, UK).
TIRF was used to detect asphaltene precipitates in the microchamber. The principle of TIRF detection is sketched in Fig. \ref{fig.2}b. Light refraction follows Snell's Law, as shown in Equation (\ref{eq1}):

\begin{equation}
n_{1}sin\theta_{1} = n_{2}sin\theta_{2}
\label{eq1}
\end{equation}
n$_{1}$ and n$_{2}$ are incident and refracted index of a given pair of media, separately, $\theta_{1}$ and $\theta_{2}$ are incident and refracted angle of light in the two media, separately. When the light propagates in a denser medium (e.g the glass substrate) and meets the interface with a less dense medium (e.g the sample on the substrate) above a critical angle, rays of light are no longer refracted but totally reflected in the denser medium. Total internal reflection (TIR) occurs, generating an evanescent field adjacent to the interface of the media. The electromagnetic wave in the evanescent field can excite the fluorophores in the sample with penetration depth around 100-200 $nm$, depending on the optical properties of the media and the incident angle.\cite{fish}

In our measurements of asphaltene precipitation, the refractive index of asphaltene (RI = 1.72) was used base on the literature\cite{buckley1998,wattana2003}, which is different to toluene (RI = 1.50)\cite{kedenburg2012} and n-pentane (RI = 1.36)\cite{kerl1995}. An excitation laser with wavelength $\lambda$ = 488 $nm$ was used to excite fluorophores and the emission wavelength detected by the detector was up to 576 $nm$. The TIRF images were taken on an 82.5 $\mu m$ $\times$ 82.5 $\mu m$ from the field of view in the area of the narrow main channel by a pco.edge sCMOS camera. Within this excitation and emission wavelength range, the asphaltene in n-pentane and toluene solution has much stronger signal than precipitated asphaltene particles, resulting in good contrast of image with the asphaltene particles shown as black particles.

For the visualization of $\beta$-alanine crystallization, the device was prefilled with a ternary mixture comprising of 6\% $\beta$-alanine (Acros organics, 99\%), 31.5\% isopropyl alcohol (Fisher Scientific, 99.9\%), and 62.5\% water by weight. Pure isopropyl alcohol was injected through the side channel at 12 $mL/h$ to drive the crystallization. The process was monitored with the microscope in bright field imaging mode through a 10$\times$ magnification objective.

\section{Results and discussion}
\label{sec:3}
\subsection{Diffusive mixing characteristics in the form of spatial and temporal solvent concentration}
\label{subsec:3.1}
\begin{figure*}
	\centering
	\includegraphics[width=\textwidth]{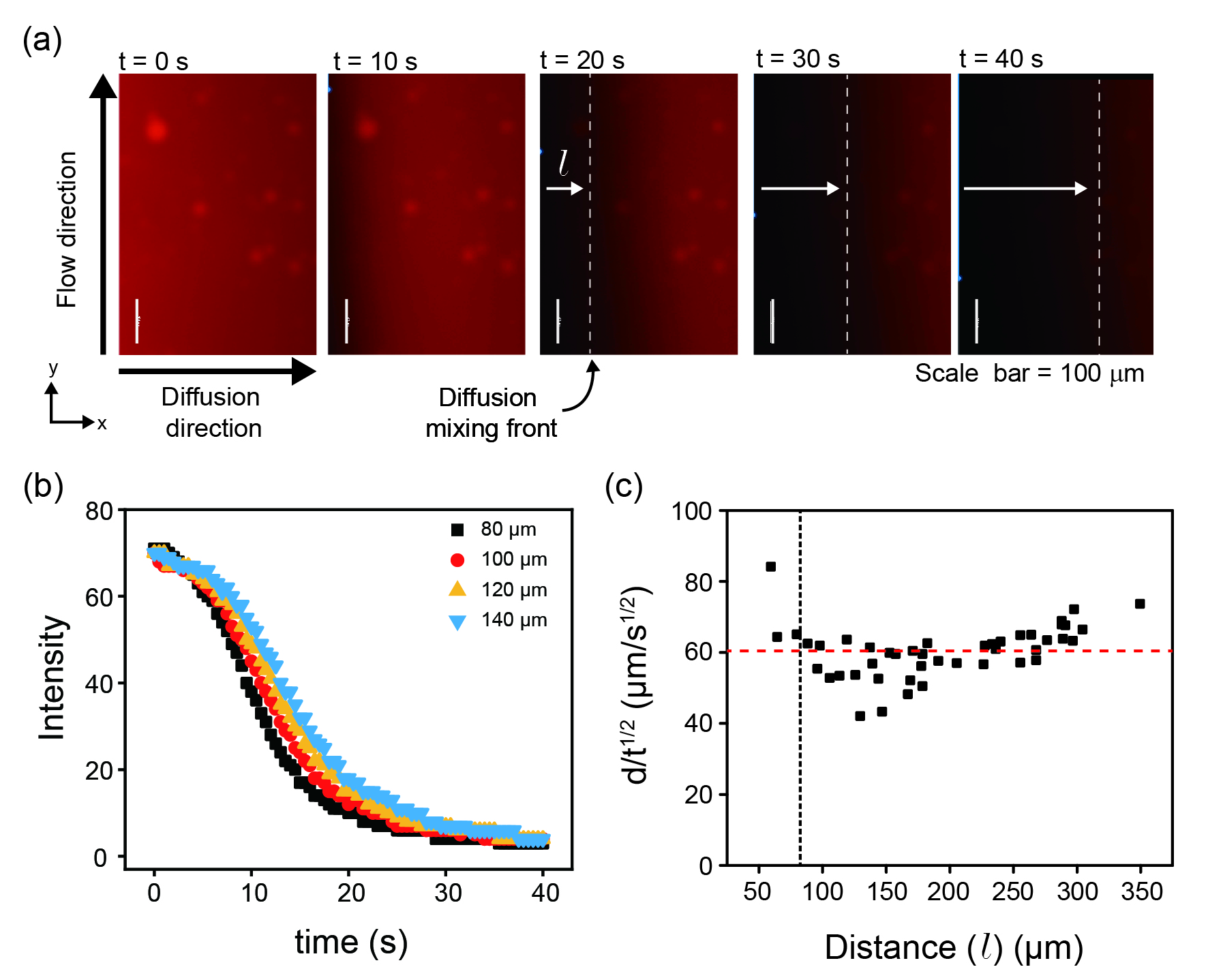}
	\caption{Fluid chamber characterization a) Snapshots of n-pentane diffuses in Nile red, toluene solution. y-axis is the flow direction, while x-axis is the diffusive mixing direction. b) Intensity versus time at different distances to the side channel. c) Front layer diffusion distance over time square root at different times. Here the front layer diffusion distance is the distance between the side channel (x = 0) and the diffusion front obtained from fluorescent images. Black dot line indicates the minimum distance required to achieve diffusive mixing (i.e. x = 80 $\mu m$). Red dashed line represents the plateaued value ($\sim$ 60 $\mu m/t^{1/2}$).}
	\label{fig.3}       
\end{figure*}

Fig. \ref{fig.3}a shows the snapshots of fluorescence images of the diffusion of n-pentane in toluene solution with Nile Red in the microchamber. At time of 0 $s$, the fluorescence signal shown as red is strongest as n-pentane just starts to reach the shallow main channel from the deep side channel. With time, n-pentane moves along the x-direction into the main channel as revealed by expansion of the dark region along x-direction in the field of view because Nile Red does not fluoresce in n-pentane within the selected excitation. After 40 $s$, the entire field of view (691.2 $\mu m$ $\times$ 491.5 $\mu m$) turns dark, suggesting that toluene is entirely displaced by n-pentane from the examined area. As n-pentane diffuses into the main channel, toluene also diffuses into the side channel. Toluene that diffuses into the side channel is pushed away by the flow in the side channel.

The fluorescence intensity at four locations is plotted as a function of time shown in Fig. \ref{fig.3}b. The locations are at 80-140 $\mu m$ away from the junction of the side channel and the main channel. The decay of the intensity (i.e. fluorescence signal vs. time) at any given location was fitted well by a sigmoid function with Equation (\ref{eq2}), derived by the same method for fluorescence in polymerase chain reaction (PCR)\cite{rutledge2004,campa2015}. As shown in Equation (\ref{eq2}), time (\textit{t}) is used to replace the number of cycles in PCR.

\begin{equation}
I = 1-\left(0.993 + \left(\frac{-1.05}{1+exp\left({\frac{t}{4.3}}-A\right)}\right)\right)
\label{eq2}
\end{equation}

\textit{I} is normalized intensity, \textit{t} is time and \textit{A} is a constant depending on the location. The delay in the intensity drop of location correlated well with the distance leading to a steady increase of 0.3 per 20 $\mu m$ distance increment for parameter A, as shown in Fig. S1, suggesting that the mixing conditions are identical at different locations in our observation area.

In Fig. \ref{fig.3}c, the front layer moving distance d normalized by the square root of time t$^{1/2}$ is plotted as a function of the distance d from the side channel. For a constant concentration source Fick's diffusion in 1-dimension scenario, the diffusion length \textit{d} and time \textit{t} follows:

\begin{equation}
\frac{d}{\sqrt t} \sim 2\sqrt D
\label{eq3}
\end{equation}

\textit{D} is diffusion coefficient, which is assumed to be constant in our system. The results of constant $\frac{d}{\sqrt t}$ suggest that mixing at a critical distance (i.e. 80 $\mu m$ to the side channel) is well described by a diffusion process. At a distance from the side channel is less than 80 $\mu m$, the mixing is not well described by diffusion, which could be caused by the influence of the side channel. On the other hand, a longer distance means a longer diffusion time, which is undesirable as the maximal fluorescence intensity is reached in a shorter duration. Therefore, our observation area is kept at 80 $\mu m$ to the side channel in the following experiments.

The duration time for n-pentane replacing toluene is about 40 $s$ over 491.5 $\mu m$ distance. Since n-pentane diffusion rate follows Equation (3), $\frac{d}{\sqrt t}$ is constant, at around 60 $\mu m/s^{1/2}$, as shown in Fig. \ref{fig.3}c. At 40 $s$, the calculated travel distance for front layer is 379 $\mu m$, which is comparable with the actual moving distance, i.e. 491.5 $\mu m$. The difference between calculated and actual values may be from the non-diffusive behaviour below the critical distance of 80 $\mu m$.

Given that n-pentane diffusion from side channel into the main channel is a one-dimension diffusion in our system, the diffusion follows:

\begin{equation}
\frac{\partial c}{\partial t} = D\frac{\partial^{2} c}{\partial x^{2}}
\label{eq4}
\end{equation}

$c$ is concentration of n-pentane, $t$ is time, $D$ is diffusivity of n-pentane in toluene and $x$ is the distance to the side channel. Solving Equation (\ref{eq4}), n-pentane concentration distribution along x-direction is:

\begin{equation}
c = c^{pen}_{B}erfc\left(\frac{x}{2\sqrt Dt}\right)
\label{eq5}
\end{equation}

For a given time, n-pentane concentration distribution along x-direction follows error function. The error function distribution is proved in Fig. S2.

\begin{figure}
	\includegraphics[width=0.5\textwidth]{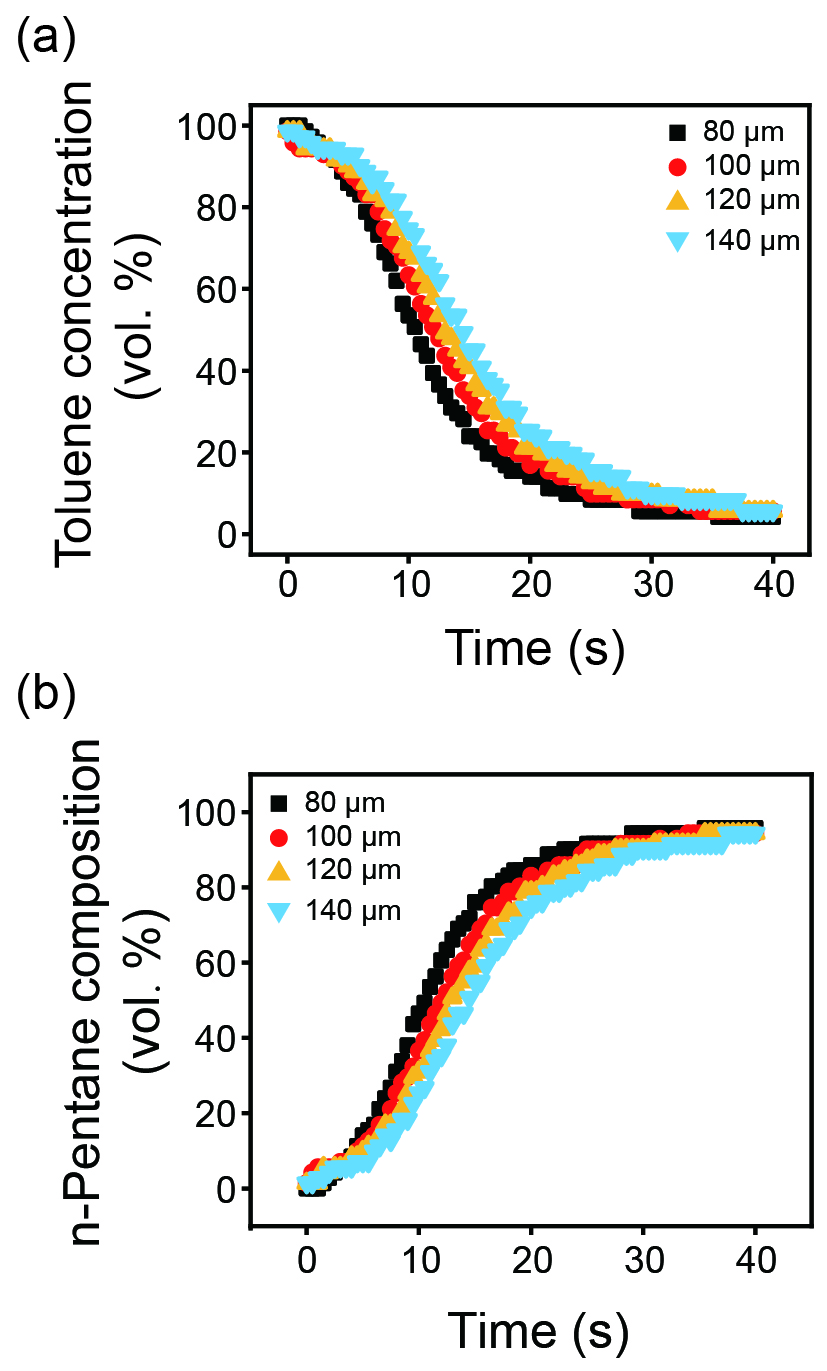}
	\caption{Change in toluene concentration with time at 80 $\mu m$, 100 $\mu m$, 120 $\mu m$ and 140 $\mu m$ of distance ($l$) to the side channel. b) Change in n-pentane concentration with time at 80 $\mu m$, 100 $\mu m$, 120 $\mu m$ and 140 $\mu m$ of distance to the side channel.}
	\label{fig.4}       
\end{figure}

Assuming the ratio of the fluorescent dye Nile Red and toluene remains constant, from the intensity with time, we can obtain r$_{solvent/toluene}$ in the diffusive mixing process. As shown in Fig. \ref{fig.4}a and b, the intensity of fluorescence is quantified to obtain the change in the chemical composition in space and in time. Toluene concentration is estimated by $c$ $\sim$ $I$. Here $c$ is toluene concentration, and $I$ is normalized fluorescence intensity. Toluene concentration decreases with diffusion time, while n-pentane concentration, reverse of the toluene concentration, increases with time. 

\subsection{Applications of the diffusive mixing device}
\label{subsec3.2}
\begin{figure*}
	\centering
	\includegraphics{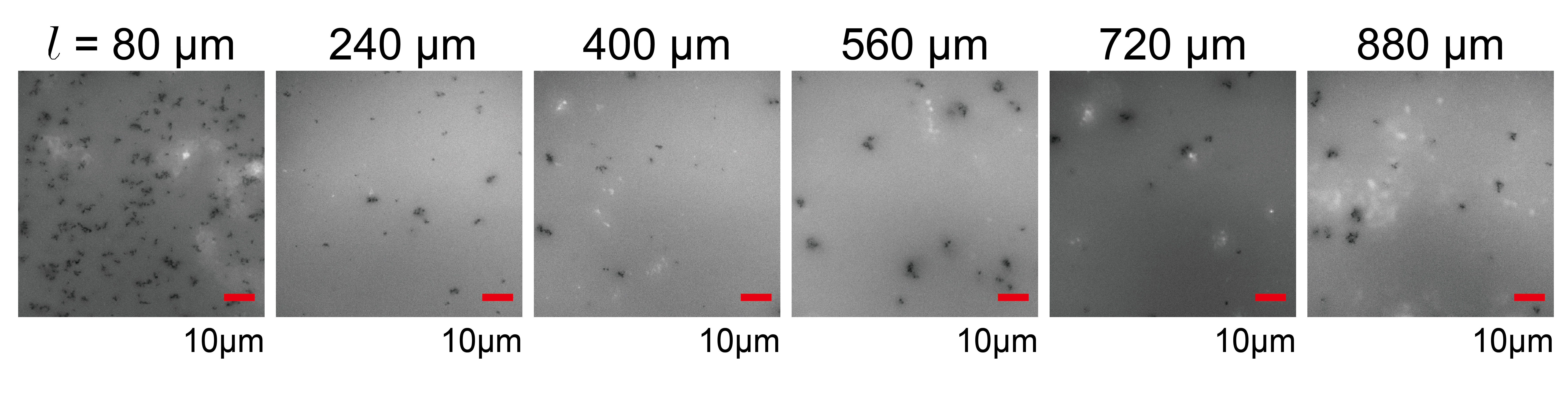}
	\caption{Total internal reflection (TIRF) microscopy images of asphaltene precipitation driven by diffusive mixing between an asphaltene solution in toluene as solution A and n-pentane as solution B. More precipitates observed at \textit{l} of 80 $\mu m$ is attributed to the difference in mixing condition near the side channel ($l$ $<$ 80 $\mu m$) and in the inner chamber ($l$ $>$ 80 $\mu m$)}
	\label{fig.5}       
\end{figure*}

The mixing device was used to observe precipitation of asphaltene from toluene by n-pentane. As shown in Fig. \ref{fig.5}, the precipitation of asphaltene can be observed in situ. Images were taken from 80 $\mu m$ which is the critical distance need to achieve diffusive mixing as also shown in Fig. \ref{fig.3}. Up to a distance of 80 $\mu m$, mixing may be influenced by the bulk flow of solution B. The high amount of precipitates observed at 80 $\mu m$ is likely due to the convective mixing. However, at longer distances, the diffusive mixing is established and the amount of precipitates remain similar with distance. With the high resolution of TIRF – i.e. 200 $nm$ – observation of individual asphaltene precipitates may also be possible.

The proposed device is not limited to the in situ asphaltene precipitation, but it can also be applied for studying dilution induced crystallization from a liquid mixture. The dilution induced crystallization occurs due to the reduced solubility of the solute in the liquid mixture by the addition of an antisolvent. Using the diffusive mixing device, we can visualize the crystallization process. As an example, we demonstrate in situ growth of $\beta$-alanine crystals by dilution of a ternary solution consisting of $\beta$-alanine, isopropyl alcohol, and water (solution A) by isopropyl alcohol (solution B). As solution B is diffused through the main channel, $\beta$-alanine becomes oversaturated and crystallizes on the surface of the channel as shown in Fig. 6. Using the device shown here, it is possible to control the crystallization by tuning the mixing between antisolvent and the liquid mixture.

\section{Conclusion}
\label{sec:4}
A method for in situ observation of precipitation and phase separation is shown here via a quasi-2D microfluidic device coupled with total internal reflection (TIRF) microscopy. The device enables controlled diffusive mixing between an asphaltene solution and a mixture of n-pentane and toluene. Using a fluorescent dye, the diffusive mixing front is determined, based on which a critical distance required to achieve fully diffusive mixing is obtained. Moreover, from the intensity profile of the fluorescent dye, the concentration profiles of n-pentane and toluene can be estimated. TIRF image shows the diffusive mixing is not influenced by asphaltene precipitation as well as the quantity of the precipitated particles. The set-up shown in this work is useful for high spatiotemporal study of precipitation dynamics for various applications including asphaltene precipitation, nanoparticle formation, crystal growth or protein precipitation.

\section{Acknowledgements}
\label{acknowledgements}
This work is supported by the Institute for Oil Sands Innovation (IOSI) (project number IOSI 2018-03) and from the Natural Science and Engineering Research Council of Canada (NSERC) – Collaborative Research and Development Grants. The authors are grateful for technical support from IOSI lab, particularly from Lisa Brandt and Brittany MacKinnon. We are also grateful for the technical support of Dr. Xuejun Sun at the Cell Imaging Facility at the Cross-Cancer Institute and Dr. Murray R. Gray in Alberta Innovates for fruitful discussions.

%
%
\section{Authors contributions}
All the authors were involved in the preparation of the manuscript.
All the authors have read and approved the final manuscript.

\newpage

\newpage

\newpage

\newpage

\newpage

\newpage
\begin{figure*}
	\centering
	\includegraphics{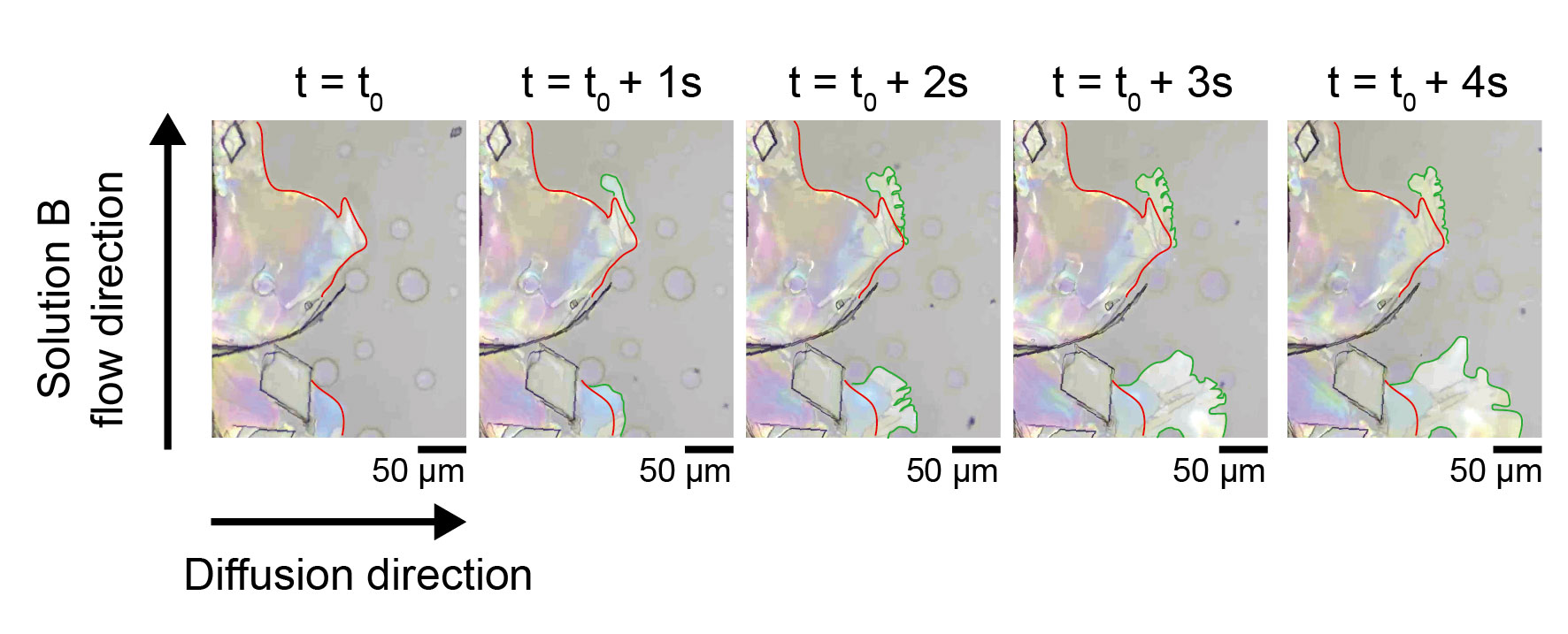}
	\caption{In situ crystallization of $\beta$-alanine using the diffusive mixing device. With time crystals --indicated in green line -- grow in the direction of diffusive mixing. The red lines indicate the boundary of stationary crytals as reference.}
\end{figure*}

\end{document}